# Encoding the Subsurface in 3D with Seismic

[1]Ben Lasscock*, [1]Altay Sansal, Alejandro Valenciano
[1]TGS

## SUMMARY

This article presents a self-supervised generative AI approach to seismic data processing and interpretation using a **M**asked **A**uto**E**ncoder (MAE) with a **Vi**sion **T**ransformer (ViT) backbone. We modified the MAE-ViT architecture to process 3D seismic mini-cubes to analyze post-stack seismic data. The MAE model can semantically categorize seismic features, demonstrated through t-SNE visualization, much like large language models (LLMs) understand text. After we fine-tune the model, its ability to interpolate seismic volumes in 3D showcases a downstream application. The study's use of an open-source dataset from the "Onward - Patch the Planet" competition ensures transparency and reproducibility of the results. The findings are significant as they represent a step towards utilizing state-of-the-art technology for seismic processing and interpretation tasks.

## INTRODUCTION

Seismic data are crucial to exploration efforts in the oil and gas industry. Despite the quantitative nature of many geophysical workflows, seismic processing and interpretation are resource-intensive and time-consuming. They present challenges due to the immense volumes of data, complexities in 3D visualization, and limited searchability. Recent advancements in artificial intelligence (AI) and machine learning (ML), particularly in supervised learning, have shown promise in automating specific processing and interpretation tasks, such as denoising and salt segmentation (Brusova et al., 2021; Warren et al., 2023 ). An essential limitation of these supervised models' is the dependence on labeled data, which means they cannot utilize all available unlabeled seismic data to improve performance.

To address this limitation, we propose a new approach to seismic processing and interpretation based on self-supervised (Gui, 2023) transformer models (Vaswani et al. 2017), inspired by the success of LLMs in processing text, introduction of image foundational models, and analogous to developments in medical imaging field (Zhou et al., 2023; Ma et al., 2024). A 2D MAE-ViT (He et al., 2021; Dosovitskiy et al., 2020) architecture was reconfigured from an open-source model (Hugging Face, 2024) to process 3D seismic inputs. The 3D model was pre-trained, self-supervised for post-stack seismic volume reconstruction, and fine-tuned for seismic interpolation. Sheng et al. (2023) similarly applied 2D MAE-ViT models to open seismic data and various downstream applications.

Our study uses an open-source dataset from the "Onward - Patch the Planet" competition (Onward, 2023) to promote reproducibility and allow for verification by the geophysical community. After we pre-train the encoder, we apply the t-SNE algorithm (van der Maaten et al., 2008; Pedregosa et al., 2011) to reduce the dimension of and then visualize the encoder's embeddings, which provided insights into its categorization capabilities and illustrated the model's interpretive power. This approach follows an LLM demonstration (OpenAI, 2023) showing how textual embeddings organize the model's parameter space according to context and meaning. Organized embeddings from the encoder make semantic search capabilities possible on seismic data. To underscore the practical relevance of this work and its potential to contribute to seismic data analysis significantly, we fine-tune the model for seismic interpolation. Our model's accuracy in interpolating post-stack seismic volumes is demonstrated, and Onward evaluated its strong performance in the competitive framework.

## METHODOLOGY

### Dataset

The training uses 50 field and 500 synthetic (300, 300, 1259) sized post-stack seismic volumes from the Patch the Planet competition (Onward, 2023). We used $256^3$ overlapping mini cubes sampled from these volumes for training tasks, applying inline/crossline flip and scaling augmentations.

### Model Architecture

In our study, instead of applying the Hugging Face (Hugging Face, 2024) MAE-ViT model on 2D images extracted from 3D seismic data, we extended the model to work on 3D seismic volumes directly. To enable this, we modified the MAE-ViT encoder and decoder. This includes changing the positional encodings to handle 3D patches and the patchify/unpatchify logic. For fast data loading and augmentation, we built a 3D data pipeline to serve 3D mini cubes using MDIO (Sansal et al., 2023). The flexibility of the transformer architecture allowed us to modify only the tokenization and positional embedding of images rather than changing the lower-level transformer layers. Once the mini cubes are patchified into 3D and are embedded, a regular transformer architecture can be used.

A crucial element of the MAE-ViT architecture is to use a random masking strategy of 3D patches in the ViT encoder and a decoder that learns to reconstruct masked portions of the seismic image. This enables the encoder to learn dense and semantically meaningful embeddings (learned representations). The masking strategy also significantly reduces the memory requirement on the MAE-ViT pre-training because the large encoder's multi-head attention only gets applied to visible tokens. The model optimizes for normalized pixel-wise L2-loss. Sine/Cosine functions created the positional encodings (Vaswani et al., 2017). In pretraining, the encoder had a hidden layer size of 768 units, comprising 12 transformer layers, employed 12 attention heads per layer, and an intermediate size of 3072 for the feedforward networks. The decoder was smaller, with 528, 8, 16, and 2048 hidden layer sizes, transformer layers, attention heads, and intermediate size of the feed-forward network. The model has approximately 120 million trainable parameters.

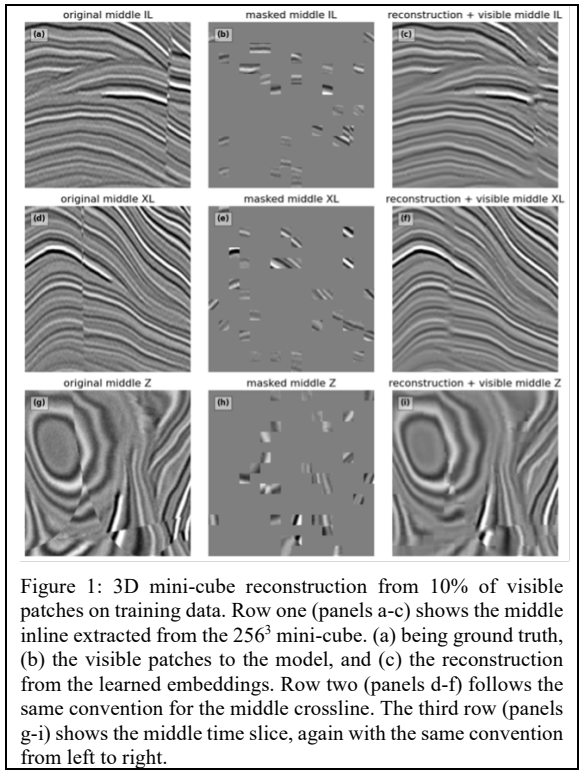

Figure 1: 3D mini-cube reconstruction from 10% of visible patches on training data. Row one (panels a-c) shows the middle inline extracted from the $256^3$ mini-cube. (a) being ground truth, (b) the visible patches to the model, and (c) the reconstruction from the learned embeddings. Row two (panels d-f) follows the same convention for the middle crossline. The third row (panels g-i) shows the middle time slice, again with the same convention from left to right.

The MAE pre-training technique randomly samples 10% of the mini-cubes with data augmentations, which is then used to reconstruct 90% of the mini-cube. We show an example of this with 2D slices through a 3D mini-cube in Figure 1, (left) the input, (center) 10% of the 3D mini-cube sampled for training, and (right) the reconstructed image. The model was trained for six weeks with an NVIDIA RTX 4090 (24GB VRAM) until achieving an acceptable level of convergence. We find that with only 10% of the original image, fine details such as faults and onlaps have become visible in the reconstruction, indicating that the learning algorithm is extracting meaningful representations.

To explore the embedding space created by the seismic encoder network, we randomly sampled 1,000 3D mini cubes from the open data library and ran them through the encoder. We use the t-SNE algorithm to create a 2D projection for visualization using a cosine metric for separation (one minus the cosine similarity) from this collection of embeddings. We then explored this parameter space using an interactive web application.

A version of the model was fine-tuned from the pre-trained network to interpolate 3D contiguous blocks of missing traces. The interpolation task is very similar to the 3D reconstruction task. The trained model used the same network architecture in pre-training with weights initialized from the pre-trained model. In this case, the encoder weights were frozen, and the decoder weights were refined. The decoder has only 30 million trainable parameters; hence, the memory requirement is significantly smaller.

Because we limited our analysis to the datasets provided by the competition, the fine-tuning reused the same augmented datasets used in training the encoder. We added custom logic to mask (full trace length) swaths of seismic data in the inline/crossline dimensions used in training. The fine-tuned model was trained for two weeks using the same hardware. It is worth noting that the size of the missing data was 29% of the input mini cube, which increased the memory requirement by approximately 7x. However, because the weights of the encoder are frozen, the overall memory requirements were lower, and it was not an issue when fine-tuning the decoder.

**EXAMPLES**

**Exploring the Embedding Space**

In Figure 2, we show an example of a mini-cube's 3D patch embeddings, which can be thought of as a projection of the input 3D patches into the high-dimensional feature space. We apply the t-SNE algorithm to 1,000 randomly sampled mini-cube embeddings to make sense of this. The t-SNE is a dimensionality reduction algorithm that preserves local relationships and projects the data to 2D space. In this example, similarity is defined by a cosine metric. Cosine similarity is a standard metric used in LLMs; an analogous example applied to text-based search and classification is shown here (OpenAI 2023). In Figure 3, we show the visualization of t-SNE output. Each point in the image is associated with a mini cube. In Figure 4, we select 2D middle-inline views related to a collection of mini-cubes sampled at different locations. Samples near the synthetic water bottom are grouped into a cluster, shown in examples (a) and (b). The field data is clustered separately from the synthetic data, shown in examples (c) and (d). The remainder of the space is less structured; however, neighboring patches are qualitatively similar. Examples (e) and (f) show mini-cubes from neighboring points in the t-SNE image, which are high amplitude continuous seismic with some faulting. This indicates that the encoder is learning a representation of the embedding space that organizes the seismic data in a way that is discriminated using cosine-similarity, a prerequisite for developing search functionality from the pre-trained model.

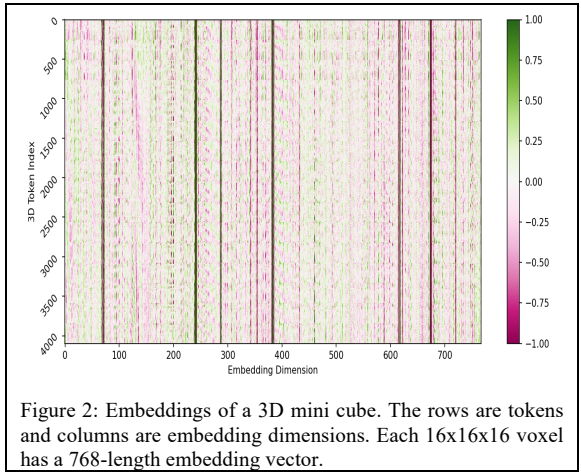

Figure 2: Embeddings of a 3D mini cube. The rows are tokens and columns are embedding dimensions. Each 16x16x16 voxel has a 768-length embedding vector.

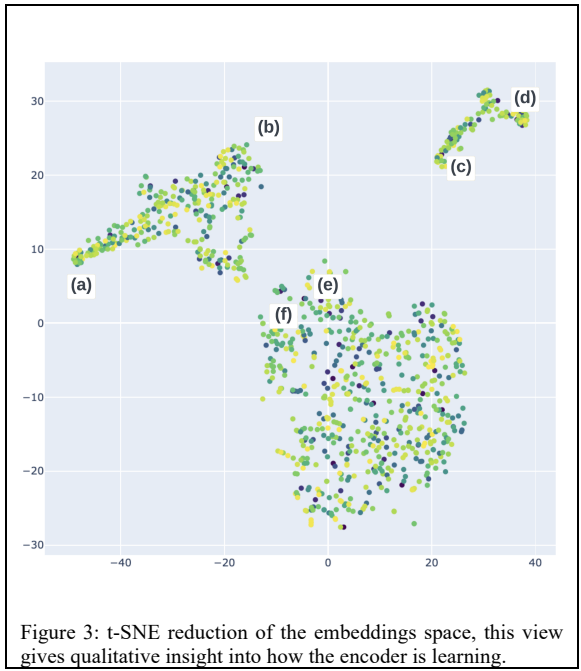

Figure 3: t-SNE reduction of the embeddings space, this view gives qualitative insight into how the encoder is learning.

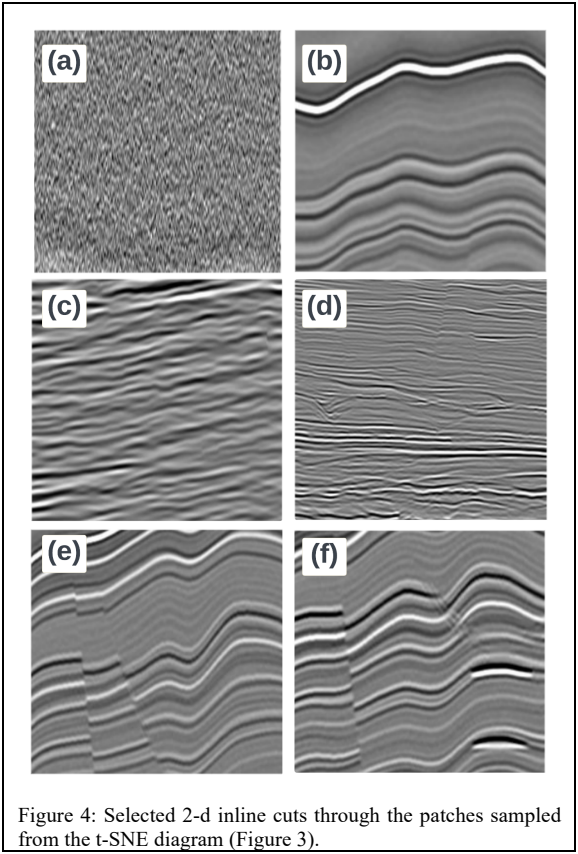

Figure 4: Selected 2-d inline cuts through the patches sampled from the t-SNE diagram (Figure 3).

### Interpolation

In this example, a model is trained to solve the "Onward - Patch the Planet" challenge. This challenge involved predicting a swath of missing data in the inline/crossline directions.

A custom model was fine-tuned for this task. Figure 5 shows an example of the results: (left) ground truth data, (center) input to the model, with a swath of data missing, and (right) a reconstruction made by the fine-tuned model. Any idiosyncratic noise is removed in the reconstruction as the model has yet to learn to generate it. The interpolation model achieved a structural similarity score 0.6025, evaluated by Onward, on a sample of three inlines and three crosslines derived from a collection of 15 held-out seismic datasets. Figure 6 shows a schematic of how the three inlines and three crosslines were sampled for scoring.

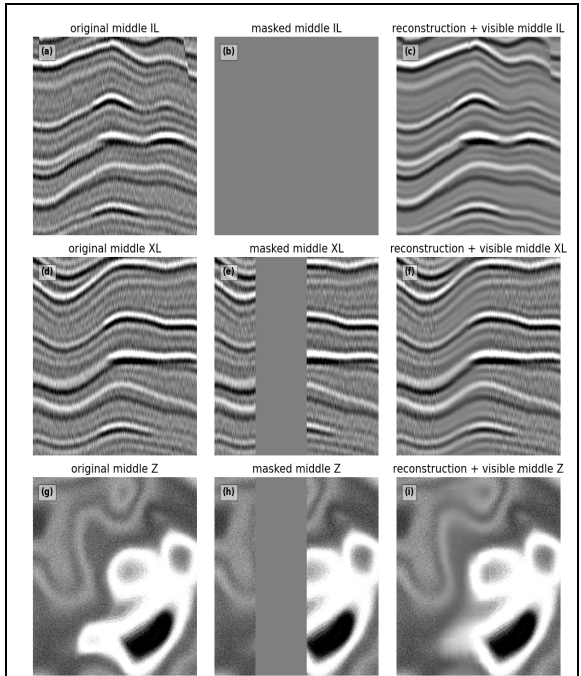

Figure 5: (left) Inline, crossline, depth slices (center) masked input reconstructed image (right).

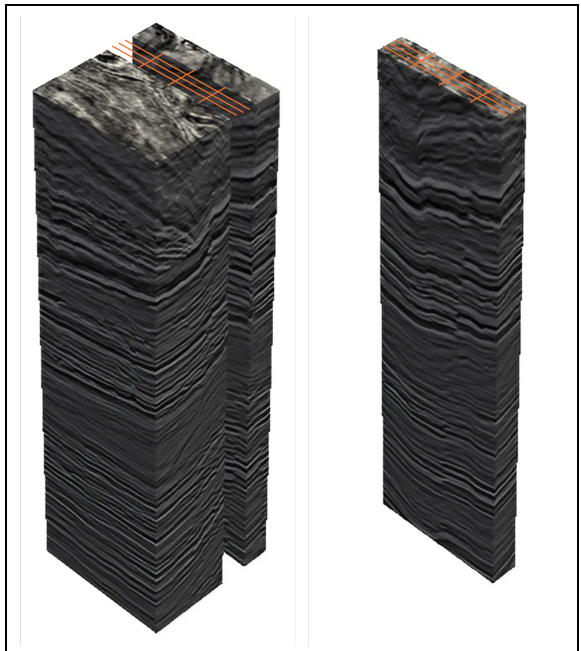

Figure 6: Schematic view of showing how the three inlines and three crosslines were sampled for scoring the Onward challenge.

## CONCLUSIONS

This study showcases how an architecture based on transformers, specifically a Masked Autoencoder, with a Vision Transformer backbone, can be effective in analyzing seismic data. By utilizing a self-supervised approach, the model can efficiently learn and reconstruct seismic features without the need for labeled data. This highlights its potential for broad applications in geophysics. The analysis of the model's feature space and its successful application in seismic data interpolation also emphasize its feature extraction capabilities and its usefulness in data processing. This research underscores the promise of transformer models in seismic analysis, setting a foundation for their expanded use and further development in the field.

## ACKNOWLEDGMENTS

The authors would like to acknowledge Onward and the "Patch the Planet" competition for providing the open-source datasets used in this study under a CC-BY 4.0 license.